# Confinement-controlled pathways to complex skyrmionic textures in Co/W/Pt multilayers

Y. Al Sadi[1], R. Sbiaa[1*], W. Al Saidi[2], M. Souier[1], G. Lezier[3], O. Marbouh[3], M.T.Z. Myint[1], Y. Dusch[2], S. Al Harthi[1], A. Talbi[2], N. Tiercelin[2] and S. N. Piramanayagam[4]

[1]Department of Physics, College of Science, Sultan Qaboos University, P.O. Box 36, PC 123, Muscat, Oman

[2]Department of Physics, College of Applied Sciences and Pharmacy, University of Technology and Applied Sciences - Muscat (UTAS-Muscat), Oman.

[3]University of Lille, CNRS, Centrale Lille, Université Polytechnique Hauts-de-France, UMR 8520 - IEMN, 59000 Lille, France

[4]School of Physical and Mathematical Sciences, Nanyang Technological University, 21 Nanyang Link, Singapore, 637371

**Magnetic skyrmions and higher-order topological spin textures offer rich opportunities for multi-level information encoding, yet their deterministic stabilization and transformation under geometric confinement at room temperature remain poorly understood. Here, we demonstrate that geometric confinement acts as a robust and universal control parameter that governs a hierarchical transformation pathway of chiral spin textures in Pt/Co/W multilayer micro-tracks. As the confinement increases, extended labyrinth domains fragment into isolated skyrmions, followed by the systematic suppression of skyrmion pairs and the preferential stabilization of compact higher-order textures. We find that confinement strongly enhances the formation of skyrmioniums via recombination and promotes their subsequent evolution into uniform skyrmion bags by capturing additional skyrmions. Statistical analysis reveals a confinement-driven redistribution of topological populations, with skyrmion bags emerging as the dominant state in the narrowest tracks. Supported by micromagnetic simulations, our results establish geometric confinement as a deterministic selector of complex topological textures and reveal a previously unexplored route for engineering higher-order skyrmionic states at room temperature. These findings provide a scalable materials strategy for multistate skyrmion-based spintronic and memory architectures.**

* Corresponding author: rachid@squ.edu.om

Magnetic skyrmions and their higher-order topological structures have received substantial attention due to their rich topological physics and technological features. Their nanoscale size, efficient current-driven motion, and topological stability[1–3] make skyrmions promising topological quasiparticles for next-generation spintronic and memory devices[4–6]. Skyrmions are nontrivial topological solitons with particle-like behavior, which were found experimentally in a wide range of magnetic materials, including bulk crystals[1,7] and sputtered thin films[8,9]. Magnetic textures other than skyrmions, such as skyrmioniums and skyrmion bags, have also been reported[10–16]. Recombination of skyrmion pairs when they have opposite topological charges results in a composite spin texture, a skyrmionium[17–20], while trapping an additional skyrmion inside a skyrmionium nucleates a skyrmion bag[11,16,21]. These topological solitons are robust, localized in space, and cannot transform into other magnetic objects of different topological charge without altering the continuity of the magnetization field[22]. The formation of these magnetic quasi-particles in chiral magnets and structured multilayers is due to the broken inversion symmetry, which induces the Dzyaloshinskii–Moriya interaction (DMI). The nucleation and stability of magnetic skyrmions are governed not only by intrinsic conditions but also by external agents like the perpendicular magnetic field[23,24], geometric confinement[25–29], and even local magnetic perturbations from a scanning probe tip[30–33]. However, the evolution of hybrid spin textures, and particularly the transformation pathway from labyrinth domains to higher-order skyrmion bags under micron-scale confinement, remains insufficiently understood. Addressing this question is essential for skyrmion-based technology, especially in racetrack-type memory, which requires the magnetic objects to be confined and stabilized in a narrow geometry.

In this letter, we employ magnetic force microscopy (MFM) to investigate the field-dependent evolution of diverse topological structures in micro-tracks of (Pt/Co/W) multilayers at room temperature. The multilayers were engineered with asymmetric stacking order and repeated interfaces to enhance the interfacial Dzyaloshinskii-Moriya interaction (DMI) and the perpendicular magnetic anisotropy, allowing for stable skyrmionic textures at room temperature. We directly observe skyrmion pairs, a $Sk^\uparrow$ with the spins directed upward at the center, and a $Sk^\downarrow$ with the spins pointed downward at the center, induced by a weak perturbation of the trivial domain state by the MFM tip. However, the skyrmion pairs get suppressed with the increase in confinement of the device. In contrast, the population of skyrmioniums increases, indicating that a recombination pathway is favored by the system. Our measurements also reveal that the geometric restriction promotes the nucleation of skyrmion bags when the skyrmioniums capture neighboring skyrmions. Our findings show that geometric confinement acts as a deterministic control parameter that selects and stabilizes higher-order skyrmionic textures through a sequential topological transformation pathway at room temperature. This reveals a previously unknown confinement-driven topological transition from simple magnetic domains to skyrmion pairs to skyrmionium and complex higher-order spin configurations. In contrast to the recent study by Kern *et al.*, which reported the formation of higher-order skyrmionic textures by using localized helium-ion irradiation to position the nanometer-scale anisotropy-engineered defects introduced in a ferromagnetic multilayer system. The study establishes the deterministic control parameter for engineering higher-order skyrmionic states[16].

The deposited sample consists of ten repetitions of (Pt/Co/W) multilayers, created by RF diode sputtering under an argon atmosphere in the $5\times10^{-3}$ to $10\times10^{-3}$ mbar range after pumping the deposition chamber to a base vacuum of $\sim10^{-7}$ mbar. The sample was patterned into micro-tracks measuring 200 μm in length, each with a unique width, using a lift-off method with a resist layer mask defined by photolithography before the sputtering step. A schematic representation of the (Pt/Co/W) multilayer structure is shown in Fig. 1(a), emphasizing the layered architecture and interfacial interactions among various constituents. The structural layers were sequenced in a particular arrangement to break structural inversion symmetry by placing Co between heavy metals with opposite spin–orbit coupling, inducing a strong interfacial DMI. The (Pt/Co/W) stack was repeated 10 times to increase the total number of interfaces, thereby increasing interfacial anisotropy and stabilizing out-of-plane magnetization in the full multilayer stack.

Magnetometer measurements carried out in the out-of-plane orientation of the deposited sample exhibit a sheared hysteresis loop and provides insight into its magnetic behavior [Fig. 1(b)]. The tilted hysteresis indicates a gradual magnetization reversal process, rather than sudden switching. The observed asymmetry originates from unequal interfacial Dzyaloshinskii–Moriya interaction (DMI) at the two interfaces, leading to distinct nucleation and annihilation fields for topological spin structures. Ultimately, these asymmetric and sheared characteristics signify the continuous evolution of magnetic moments throughout the reversal process.

A more detailed inspection of the sample using scanning electron microscopy (SEM) is shown in Fig.1(c), indicating device widths of 50, 20, and 10 μm, arranged from bottom to top.

Magnetic force microscopy (MFM) was conducted in non-contact mode at room temperature. The basic setup and operational mechanism are illustrated in a schematic diagram [Fig.1(d)]. The sample was subjected to an out-of-plane magnetic field from a permanent magnet of approximately 20 and 50 mT. The measurements were made using a low-moment CoCr (MESP-LM-V2) magnetic tip to minimize the tip-induced perturbation of magnetic structures. The system employs a frequency modulation FM-AFM operating under ultra-high vacuum. The MFM signal is obtained in a single-pass and non-contact mode. This method is different from the conventional double-pass techniques. The use of UHV and FM-MFM in non-contact mode enables stable, high-resolution MFM imaging required to resolve nanoscale magnetic textures[34]. The plane-subtract technique is used to enhance and verify that the measurements are purely magnetic by increasing the tip-sample distance to a point where other sample forces are negligible.

The magnetic configuration at remanence state for a 50 μm device indicates a well-defined labyrinth domain [Fig. 2(a)], which is typical of systems with strong perpendicular anisotropy[35–37]. When an out-of-plane magnetic field of 20 mT is applied, the bright regions within the labyrinth structure expand [Fig. 2(b)], indicating the expected alignment of the magnetic domains with the direction of the external magnetic field.

As the magnetic field increases further toward 50 mT, the maze-like domains continuously disintegrate, and isolated circular domains emerge, signifying the nucleation of skyrmions coexisting with the surviving labyrinth patterns. The formation of these nanoscale objects near the approach to magnetic saturation, where the labyrinth network splits into topologically protected nanoscale structures, is in line with previous studies[30–33]. A key finding is that the observed magnetic structures extended beyond the anticipated skyrmions. A skyrmionium with its characteristic donut-like shape that coexists with the skyrmion pair, consistent with the earlier reports suggesting that skyrmioniums can be formed from paired skyrmionic states[19]. A destructive magnetic domain next to skyrmions (shown by white arrow) and a nearby $Sk^{\uparrow}$ and $Sk^{\downarrow}$ (white dashed circle) [Fig. 2(c)], verifying the presence of mixed topological excitations.

Additionally, isolated skyrmions can also be formed due to the magnetic tip effects, which aligns with our observation of the fragmentation of dark magnetic domains into $Sk^{\downarrow}$, [blue arrow in Fig. 2(c)] and the dotted white circle [Fig. 3(a) (forward scan) and 3(b) (backward scan)] in the 20 μm microwire. Remarkably, Fig. 3(b) demonstrates a tip-induced skyrmion pair (dotted blue rectangle) nucleating from a suppressed magnetic domain during backward scan. The fragmentation of the magnetic domain and the survival of skyrmionium [Fig. 3(c) (forward scan) and 3(d) (backward scan)] at this particular magnetic field indicate the difference between the fine-tuned trivial state and protected nontrivial topological structures. The above results were frequently seen in various portions of the sample, demonstrating that the repetitive imaging enhances the creation of skyrmion pairs, which act as fundamental building blocks for the more complex magnetic spin textures identified in this study, while the topological textures remain relatively stable against the tip destruction.

These magnetic domains at remanence and at 20 mT were consistently observed across all micro-devices. However, at 50 mT, the point at which topological textures begin to form, the narrower devices exhibit a different behavior, suggesting that geometric confinement dramatically alters the stability and evolution of the emergent skyrmionic states.

For the device with 20 µm width, magnetic domains with reduced density coexist with a mixture of magnetic structures, as we observed in repeated measurements at various locations on the device. In this intermediate confinement device, compact multi-skyrmion states such as skyrmion bags emerge, recording a pronounced transition toward a new complex higher-order topological structure. In Fig. 4(a), the skyrmion bag (indicated by the red arrow) has a topological charge of $Q = 1$. Adjacent to it, a composite topological texture (indicated by the blue arrow), which is a combination of $Sk^\uparrow$ and $Sk^\downarrow$ with another skyrmion at the boundary (indicated by the white arrow). In Fig. 4(b), skyrmioniums, skyrmion pairs, and skyrmion bags show simultaneous coexistence, highlighting that multiple and composite topological states can be stabilized within the same field and confinement conditions. Of particular importance, individual skyrmions often appear at the boundaries of both skyrmioniums and skyrmion bags, suggesting that boundary-mediated interactions play a role in the internal arrangement and stabilization of these textures. The dashed blue squares in Fig. 4(b) further show skyrmion bags of different sizes, each containing two internal skyrmions, indicating the ability of the system to stabilize composite textures consisting of an outer domain wall enclosing internal skyrmions. These findings underscore that 20 µm geometric confinement further suppresses the magnetic domain and promotes the emergence and stability of various chiral magnetic states.

For device with 10 µm width, the evolution of magnetic textures under confinement is remarkable compared to the cases of 50 µm and 20 µm. As confirmed by different scanning locations, magnetic domains are fully suppressed and no longer coexist with the emergent topological textures, indicating the continuous fragmentation into skyrmions as the confinement intensifies, which has been confirmed by several theoretical and experimental studies[26,28]. As observed in the previous devices, the pair of skyrmions still coexist in this micro-device [Fig. 4(c and d)]. Markedly, under this geometric confinement, the skyrmion bags become clearer, more symmetric, and more uniform than in previous devices with larger widths, reflecting edge-driven evolution. Also, the number of these textures other than simple skyrmions is noticeably larger, confirming their high possibility. A fascinating MFM image of two skyrmioniums (2 and 4) and three skyrmion bags (1, 3, and 5) are shown in Fig. 4(c), demonstrating the favored spin configurations at this confinement and field conditions. The bags are larger in size to accommodate more than one skyrmion, which is most frequently observed near the boundaries, suggesting the possibility of complex interaction and the ability to host multiple internal skyrmions. Additionally, it is crucial to highlight the observation of an interesting magnetic feature as indicated in Fig. 4(b) by the number by 3, underscoring the wide range of skyrmionic states that emerge at the narrower device.

Overall, the above results demonstrate that the 50 mT magnetic field is within the range in which the tip stray field induces $Sk^{\uparrow}$ and $Sk^{\downarrow}$ pair and the confinement drive the evolution of higher-order topological structures [Fig. 4(e)].

By repeating the measurements at multiple locations across each micro-track, we constructed a statistical profile of the previously discussed complex topological structures, which shows that the relative populations are confinement-determined. In Fig. 5(a), the geometric confinement alters the density of the emergent states, demonstrating the increasing population of skyrmioniums and skyrmion bags while skyrmion pairs progressively decline. In narrow tracks, the degrees of freedom for the spatially separated $Sk^{\uparrow}$ and $Sk^{\downarrow}$ to coexist are reduced by the two cores interaction and the edge distortion, in contrast to skyrmioniums, a spatially compact topologically neutral structure generated by two skyrmions with opposite topological charges. This suggests that the recombination pathway of $Sk^{\uparrow}$ and $Sk^{\downarrow}$ is more favorable, resulting in the formation of skyrmioniums. Additionally, the increasing density of both mutually repulsive skyrmions (as previously discussed) and skyrmioniums in confined areas raises the interaction energy of the system. To reach a lower energy state, skyrmioniums may trap the nearby skyrmions, as frequently seen near the skyrmionium's periphery in the MFM images. Consequently, as the confinement intensifies, the possibility of skyrmioniums capturing additional skyrmions increases, which favors the generation of skyrmion bags, the dominant state in the narrowest device. It is worth mentioning that the tip perturbation and the confinement are a pathway selector. Weak and localized perturbations do not destroy but instead catalyze topologically allowed transformations that are subsequently stabilized by the confinement [Fig. 5(b)].

Different scanning-height measurements were performed in this study to emphasize that the textures are purely magnetic, rather than influenced by surface topography [Fig. 5(c-e)]. At a height of 50 nm, the tip mostly senses the long-range magnetic interaction, excluding any possible topographic contribution from the sample's surface. Therefore, the contrast perseverance clearly indicates the magnetic origin of the observed textures. When the tip height was decreased to 20 nm, the gradient of magnetic force increased, and the internal cores of both skyrmionium and skyrmion bags became significantly clearer, as the resolution of MFM imaging is distance-dependent. The long-range contrast and its distance-dependent further highlight the magnetic nature of the observed topological structures throughout the study.

The micromagnetic simulation was conducted using the MuMax3 software[38], which is based on the modified Landau-Lifshitz-Gilbert (LLG) equation to model the dynamics of magnetization that is expressed as

$$\frac{d\mathbf{m}}{dt} = \frac{-\gamma}{1+\alpha^2} [\mathbf{m} \times \mathbf{H}_{eff} + \mathbf{m} \times (\mathbf{m} \times \mathbf{H}_{eff})] \quad (1)$$

where $\mathbf{m}$, $\gamma$ and $\alpha$ represent the normalized magnetization vector, the gyromagnetic ratio, and the Gilbert damping coefficient, respectively. The effective field $\mathbf{H}_{eff}$ includes contributions from the exchange, anisotropy, and external fields. The calculation was performed on a sample of $512 \times 512$ nm$^2$ divided into small cubes of 1 nm edge. The material parameters, saturation magnetization, exchange stiffness, Dzyaloshinskii-Moriya interaction strength, and the Gilbert damping parameter used in the calculation were fixed to $580 \times 10^3$ A/m, $15 \times 10^{-12}$ J/m, 3.0 mJ/m$^2$, $15 \times 10^{-12}$ J/m and 0.1, respectively. The magnetization is oriented along the out-of-plane direction with a uniaxial anisotropy constant of $0.6 \times 10^6$ J/m$^3$. The evolution of the skyrmion pair is shown by the $z$-component of magnetization [Fig. 6(a-e)] taken at $t = 0$, 1, 2, 3, and 3.7 ns, respectively. We observed a series of transformations with time, starting with an initial separation distance between their cores of 40 nm. Differentiable up skyrmion core Sk$^\uparrow$ and down skyrmion core Sk$^\downarrow$ are seen at $t = 0$ ns but as time elapses, the cores become closer until both topological textures disappear. The opposite topological charges and magnetic configurations of the Sk$^\uparrow$ and Sk$^\downarrow$ cause them to annihilate each other. When the distance between them is small, a skyrmion pair approaches one another and, because of their opposite topological charges, undergoes an annihilation process that minimizes the system's energy, as has been reported[19]. Notably, the separation distance between the skyrmion pair cores affects the stability of the magnetic configuration. The annihilation of the skyrmion pair occurs at small distances. On the other hand, a new magnetic arrangement appeared at intermediate distances where the skyrmionium could be stabilized [Fig. 6(f-j)]. Interestingly, both $Sk^{\uparrow}$ and $Sk^{\downarrow}$ are preserved as their separation is increased. Under the calculation conditions above, both the $Sk^{\uparrow}$ and $Sk^{\downarrow}$ coexist for $S > 95$ nm. The $Sk^{\uparrow}$ and $Sk^{\downarrow}$ cores are visible at $t = 0$, each distinguished by its magnetization profile. The cores of the skyrmion pairs show intricate interaction with time, as represented by the $z$-component of magnetization [Fig. 6(f-j)] taken at the initial state, 10 ps, 500 ps, and 1.5 ns, respectively. The attractive force between these two entities with opposite topological charges causes the surrounding spins to reorganize and leads to the formation of a stable skyrmionium. As discussed above, the annihilation of a pair of skyrmions or the formation of a skyrmionium depends on the initial distance between them. The time $\tau$ for the formation of the skyrmionium exhibits an exponential dependence on the $Sk^{\uparrow}$–$Sk^{\downarrow}$ separation S. Fig. 6(k) reveals that for a strong attractive force between $Sk^{\uparrow}$ and $Sk^{\downarrow}$ ($S < 80$ nm), there is an annihilation of both of them as described in Fig. 6(e). As the force is reduced (80 nm $< S <$ 95 nm), the formation of skyrmionium from a pair of $Sk^{\uparrow}$–$Sk^{\downarrow}$ becomes possible. The smaller separation leads to a faster skyrmionium formation. When the attractive force becomes very weak ($S > 95$ nm), the skyrmion pair remains stable.

In summary, our systematic MFM measurements reveal that the tip-induced perturbation and geometric confinement act jointly to nucleate and stabilize higher-order topological spin structures in (Pt/Co/W) multilayer stacks at room temperature. Our findings show that narrower devices suppress the labyrinth domains and skyrmions of opposite polarity while simultaneously increasing the prevalence of multi-skyrmionic structures, such as skyrmioniums and skyrmion bags, which emerge as energetically favored configurations under confinement. These results establish geometric confinement as a novel controllable mechanism for the emergence of higher-order spin configurations in thin film at room temperature. The stabilization of complex topological spin textures in multilayer systems at room temperature opens new opportunities for exploring experimentally their fundamental properties, including their interactions and dynamical response to spin torques and other external stimuli.

**Acknowledgments**

The authors acknowledge the support of the RENATECH French network, as well as the SPINMAT project of the France 2030 PEPR SPIN program. Part of this work was realized with the support of Centrale Lille Institute.

**Author contributions**

Y.S carried out MFM measurements, R.S. designed the structures of the films, proposed the devices and supervised the study. G.L and N.T. deposited the samples, O.M. fabricated the microdevices, S.N.P. contributed to the samples design, M.T.Z.M., Y.D, A.T and S.H. contributed to the analysis of the results. All authors discussed the results and provided inputs to the manuscript.

**Competing interests**

The authors declare no competing interests

**References**

1. Yu, X. Z. *et al.* Real-space observation of a two-dimensional skyrmion crystal. *Nature* **465**, 901–904 (2010).
2. Legrand, W. *et al.* Room-Temperature Current-Induced Generation and Motion of sub-100 nm Skyrmions. *Nano Lett.* **17**, 2703–2712 (2017).
3. Chui, C. P., Ma, F. & Zhou, Y. Geometrical and physical conditions for skyrmion stability in a nanowire. *AIP Adv.* **5**, 047141 (2015).
4. Zhang, X. *et al.* Skyrmion-skyrmion and skyrmion-edge repulsions in skyrmion-based racetrack memory. *Sci. Rep.* **5**, 7643 (2015).
5. Luo, S. & You, L. Skyrmion devices for memory and logic applications. *APL Mater.* **9**, 050901 (2021).
6. Song, K. M. *et al.* Skyrmion-based artificial synapses for neuromorphic computing. *Nat. Electron.* 148–155 (2020) doi:10.1038/s41928-020-0385-0.
7. Mühlbauer, S. *et al.* Skyrmion Lattice in a Chiral Magnet. *Science (80-. ).* **323**, 915–919 (2009).
8. Tejo, F., Riveros, A., Escrig, J., Guslienko, K. Y. & Chubykalo-Fesenko, O. Distinct magnetic field dependence of Néel skyrmion sizes in ultrathin nanodots. *Sci. Rep.* **8**,

1–10 (2018).

9. Je, S. G. *et al.* Creation of Magnetic Skyrmion Bubble Lattices by Ultrafast Laser in Ultrathin Films. *Nano Lett.* **18**, 7362–7371 (2018).

10. Zhang, X. *et al.* Control and manipulation of a magnetic skyrmionium in nanostructures. *Phys. Rev. B* **94**, 094420 (2016).

11. Foster, D. *et al.* Two-dimensional skyrmion bags in liquid crystals and ferromagnets. *Nat. Phys.* **15**, 655–659 (2019).

12. Tang, J. *et al.* Magnetic skyrmion bundles and their current-driven dynamics. *Nat. Nanotechnol.* **16**, 1086–1091 (2021).

13. Seng, B. *et al.* Direct Imaging of Chiral Domain Walls and Néel-Type Skyrmionium in Ferrimagnetic Alloys. *Adv. Funct. Mater.* **31**, 2102307 (2021).

14. Nakamura, K. & Leonov, A. O. Communicating skyrmions as the main mechanism underlying skyrmionium (meta)stability in quasi-two-dimensional chiral magnets. 1–11 (2024).

15. Jiang, A., Zhou, Y., Zhang, X. & Mochizuki, M. Transformation of a skyrmionium to a skyrmion through the thermal annihilation of the inner skyrmion. *Phys. Rev. Res.* **6**, 013229 (2024).

16. Kern, L. M. *et al.* Controlled Formation of Skyrmion Bags. *Adv. Mater.* **37**, 1–12 (2025).

17. Finazzi, M. *et al.* Laser-induced magnetic nanostructures with tunable topological properties. *Phys. Rev. Lett.* **110**, 177205 (2013).

18. Ishida, Y. & Kondo, K. Theoretical comparison between skyrmion and skyrmionium motions for spintronics applications. *Jpn. J. Appl. Phys.* **59**, SGGI04 (2020).

19. Zheng, F. *et al.* Skyrmion–antiskyrmion pair creation and annihilation in a cubic chiral magnet. *Nat. Phys.* **18**, 863–868 (2022).

20. Powalla, L. *et al.* Skyrmion and skyrmionium formation in the two-dimensional magnet Cr2Ge2Te6. *Phys. Rev. B* **108**, 214417 (2023).

21. Bo, L. *et al.* Controllable creation of skyrmion bags in a ferromagnetic nanodisk. *Phys. Rev. B* **107**, 224431 (2023).

22. Braun, H. B. Topological effects in nanomagnetism: From superparamagnetism to chiral quantum solitons. *Adv. Phys.* **61**, 1–116 (2012).

23. Romming, N., Kubetzka, A., Hanneken, C., von Bergmann, K. & Wiesendanger, R. Field-Dependent Size and Shape of Single Magnetic Skyrmions. *Phys. Rev. Lett.* **114**, 177203 (2015).

24. Yang, S. *et al.* Magnetic Field Magnitudes Needed for Skyrmion Generation in a General Perpendicularly Magnetized Film. *Nano Lett.* **22**, 8430–8436 (2022).

25. Jiang, W. *et al.* Blowing magnetic skyrmion bubbles. *Science (80-. ).* **349**, 283–286 (2015).

26. Jin, C. *et al.* Control of morphology and formation of highly geometrically confined magnetic skyrmions. *Nat. Commun.* **8**, 1–9 (2017).

27. Hou, Z. *et al.* Manipulating the Topology of Nanoscale Skyrmion Bubbles by Spatially Geometric Confinement. *ACS Nano* **13**, 922–929 (2019).

28. Twitchett-Harrison, A. C. *et al.* Confinement of Skyrmions in Nanoscale FeGe Device-like Structures. *ACS Appl. Electron. Mater.* **4**, 4427–4437 (2022).

29. Wang, X. R., Hu, X. C. & Sun, Z. Z. Topological Equivalence of Stripy States and Skyrmion Crystals. *Nano Lett.* **23**, 3954–3962 (2023).

30. Casiraghi, A. *et al.* Individual skyrmion manipulation by local magnetic field gradients. *Commun. Phys.* **2**, 145 (2019).

31. Díaz, S. A. & Arovas, D. P. Quantum Nucleation of Skyrmions in Magnetic Films by Inhomogeneous Fields. in *Memorial Volume for Shoucheng Zhang* 19–33 (WORLD SCIENTIFIC, 2021). doi:10.1142/9789811231711_0004.

32. Zelent, M. *et al.* Skyrmion Formation in Nanodisks Using Magnetic Force Microscopy Tip. *Nanomaterials* **11**, 2627 (2021).

33. Ognev, A. V. *et al.* Magnetic direct-write skyrmion nanolithography. *ACS Nano* **14**, 14960–14970 (2020).

34. Giessibl, F. J. Advances in atomic force microscopy. *Rev. Mod. Phys.* **75**, 949–983 (2003).

35. Jiang, W. *et al.* Skyrmions in magnetic multilayers. *Phys. Rep.* **704**, 1–49 (2017).

36. Jena, S. K. *et al.* Interfacial Dzyaloshinskii–Moriya interaction in the epitaxial W/Co/Pt multilayers. *Nanoscale* **13**, 7685–7693 (2021).

37. Lone, A. H. *et al.* Multilayer ferromagnetic spintronic devices for neuromorphic computing applications. *Nanoscale* **16**, 12431–12444 (2024).

38. Vansteenkiste, A. *et al.* The design and verification of MuMax3. *AIP Adv.* **4**, 107133 (2014).

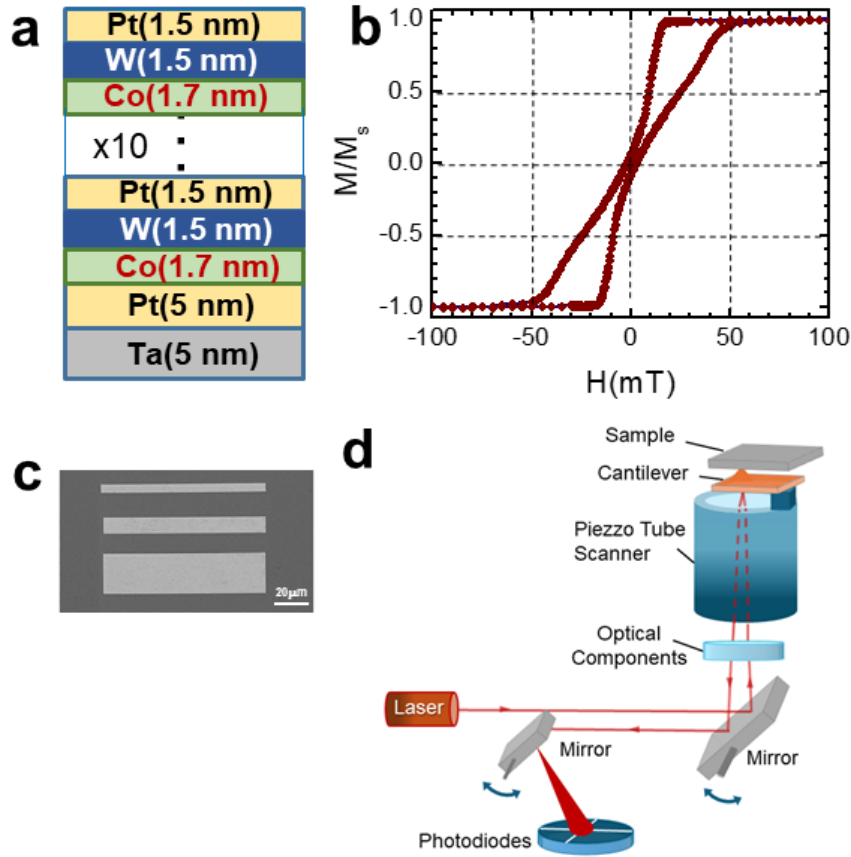

**FIG. 1│ Device structure and magnetic characterization of the (Pt/Co/W) multilayer system.** (a) Schematic of the (Pt/Co/W)$_{\times 10}$ multilayer deposited on a SiO$_2$ substrate. The asymmetric stacking and multilayer repetitions enhance the Dzyaloshinskii–Moriya interaction (DMI) and perpendicular magnetic anisotropy. (b) Magnetic hysteresis loop measured on the continuous film. The sheared loop indicates gradual magnetization rotation and a distribution of reversal fields, consistent with progressive nucleation and collapse of topological spin textures under the applied field. (c) Scanning electron microscopy images of patterned devices with widths of 50, 20 and 10 μm (from bottom to top). (d) Schematic of the magnetic force microscopy configuration used for imaging.

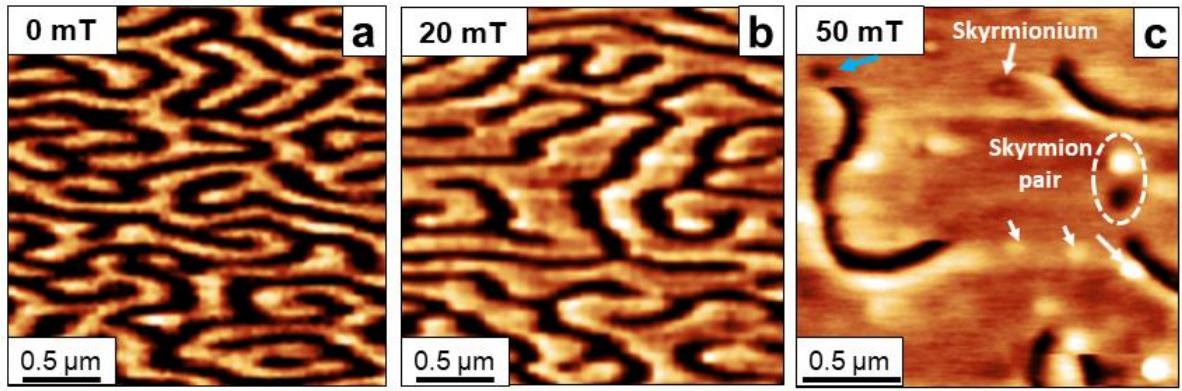

**FIG. 2│ Field-dependent evolution of magnetic domain for the device width of 50 μm.** (a) At the remanence state, the system displays a labyrinth pattern. (b) At 20 mT out-of-plane magnetic field, the bright domains expand in the direction of the field. (c) At 50 mT, the maze-like domains break up into isolated skyrmions coexisting with persistent magnetic domains. The system demonstrates remarkable magnetic configurations along with partially collapsed domains, a skyrmionium (donut-like contrast), and skyrmion pairs (dashed circle). These results show the appearance of heterogeneous topological textures in a field near the saturation.

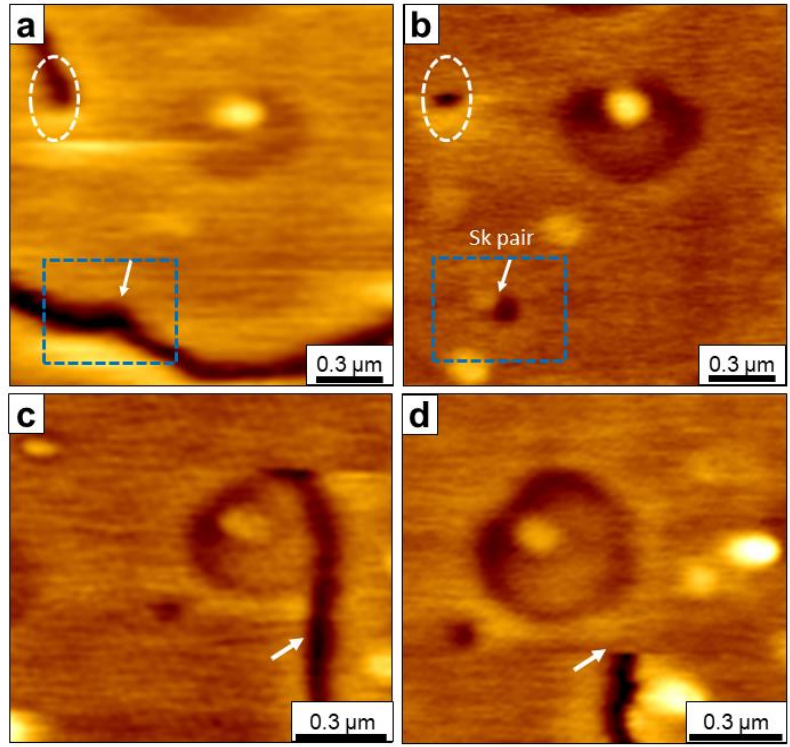

**FIG. 3 | Tip-induced formation and stability of topological spin textures.** The tip-driven effect and stability of topological textures at two distinct locations (top and bottom) under a magnetic field of 50 mT for forward scan (left) and backward scan (right). (a) and (b), the magnetic domains in the dotted white circle and dotted blue rectangle fragmented into opposite polarity skyrmions pair. c, and d, the difference between the non-topological stripe domain and stable magnetic structures under the influence of tip magnetization. Sequential scanning promotes the emergence of skyrmions pairs while preserving the topological structures.

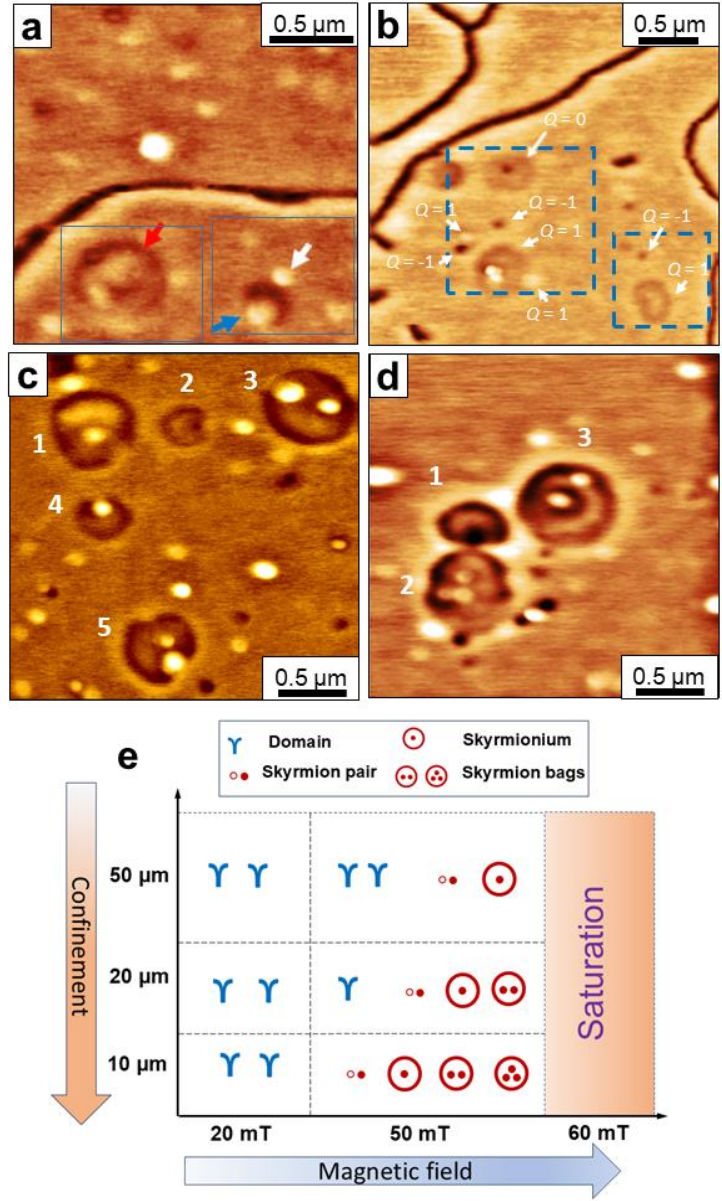

**FIG. 4 | Confinement-driven evolution of topological spin textures**. Magnetic configurations measured at 50 mT in devices with widths $w = 20$ μm (a,b) and $w = 10$ μm (c,d). Under moderate confinement ($w = 20$ μm), skyrmioniums, skyrmion pairs and skyrmion bags coexist with residual magnetic domains. (a) A skyrmion bag (red arrow) surrounded by skyrmions of opposite polarity (blue arrow) and an additional peripheral skyrmion (white arrow). (b) Skyrmioniums and skyrmion bags within the dashed blue region. With stronger confinement ($w = 10$ μm), magnetic domains are suppressed and discrete Sk↑ and Sk↓ textures dominate. (c) A symmetric skyrmion bag with topological charge $Q = 1$ (labels 3 and 5) together with a nearby skyrmion (label 1), suggesting multi-skyrmion confinement within a closed domain wall. (d) Coexistence of a skyrmion bag and a skyrmion pair (label 2), a skyrmionium (label 1), and a complex hybrid texture (label 3). (e) Phase diagram extracted from MFM imaging showing the evolution of magnetic states with lateral confinement and applied field: magnetic domains dominate below 50 mT, whereas at 50 mT diverse topological textures emerge and are progressively stabilized with increasing confinement.

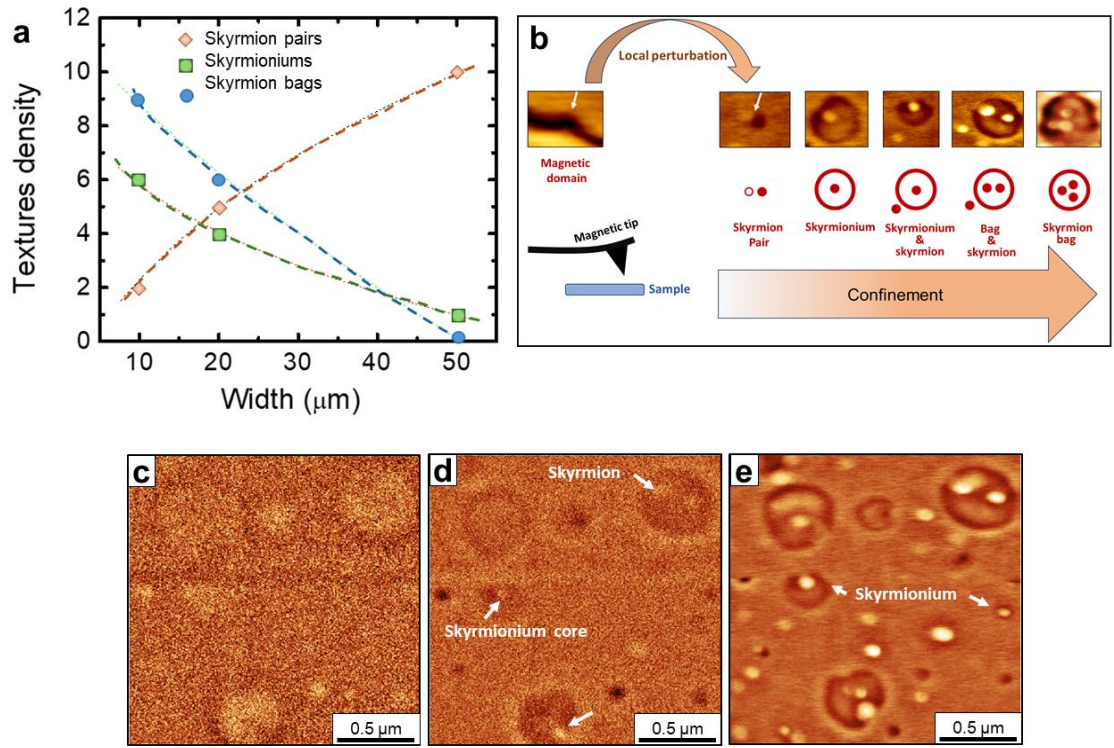

**FIG. 5 | Confinement-driven evolution and height-resolved imaging of topological spin textures.** (a) Texture density measured at six locations on the same device, demonstrating statistical reproducibility of the magnetic states. (b) Schematic phase diagram showing the evolution of topological spin textures with increasing geometric confinement. Tip-induced perturbations first generate metastable skyrmion pairs from a trivial domain state. Increasing confinement promotes recombination of skyrmion pairs into skyrmioniums, which under stronger restriction capture neighbouring skyrmions to form composite skyrmion bags. (c–e) Height-resolved magnetic force microscopy (MFM) imaging of representative textures acquired at lift heights of 50 nm, 20 nm, and 10 nm, respectively.

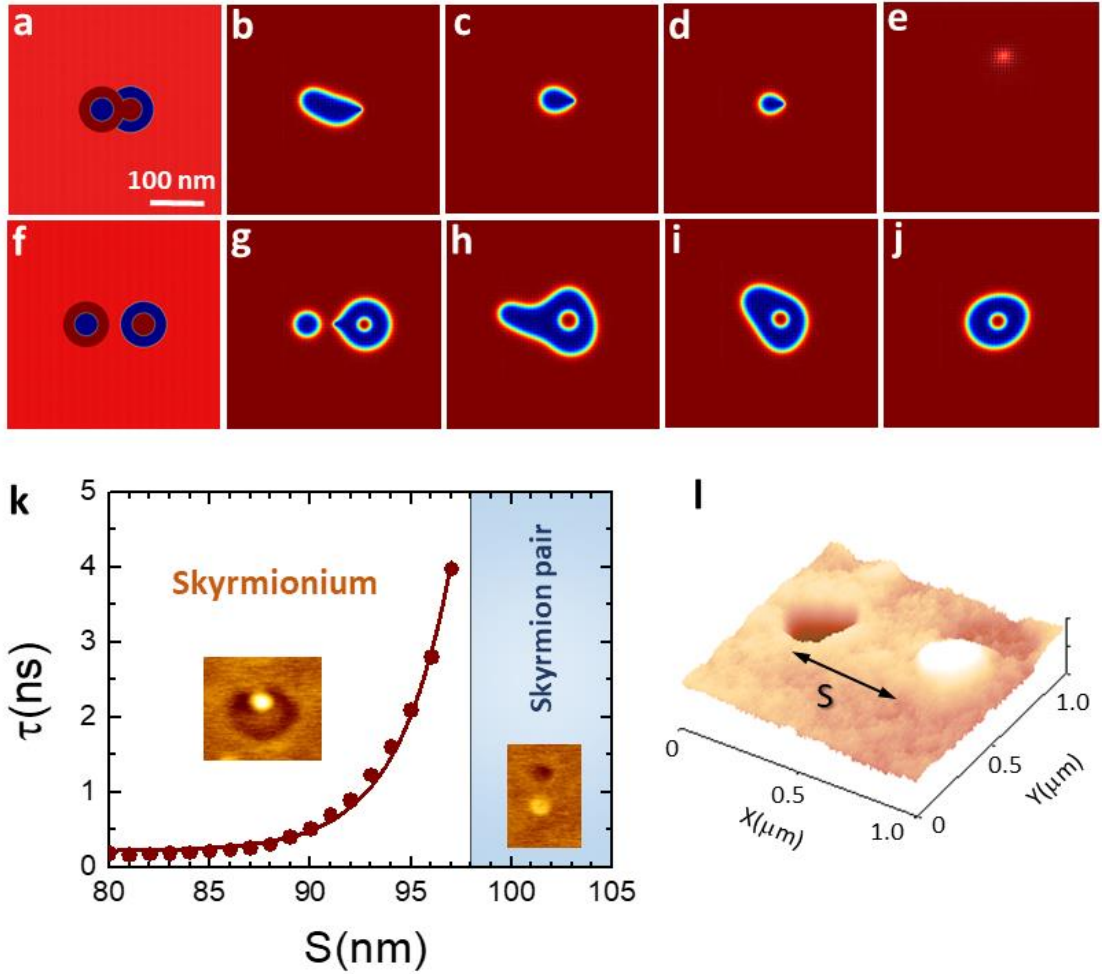

**FIG. 6 | Separation-dependent dynamics of skyrmion pairs: annihilation versus skyrmionium formation.** (a–j) Time-resolved evolution of the *z*-component of magnetization for two initial core–core separations. (a–e) For 40 nm, the skyrmion pair collapses and mutually annihilates at *t* = 0, 1, 2, 3 and 3.7 ns. (f–j) For 80 nm, the pair recombines to form a skyrmionium at *t* = 0, 10 ps, 500 ps and 1.5 ns. (k) Skyrmionium formation time as a function of the initial pair separation and (l) MFM image of the initial skyrmion pair prior to skyrmionium formation.